# Exploring the third dimension in quantum confinement of surface electrons

Lu Lyu[1,2,*], Tobias Eul[1,2], Wei Yao[1], Jin Xiao[3], Mostafa Ashoush[4], Zakaria M. Abd El-Fattah[4,5], Ignacio Piquero-Zulaica[6,7], Johannes V. Barth[8], Martin Aeschlimann[1], Benjamin Stadtmüller[1,2,*]

[1]Department of Physics and Research Center OPTIMAS, Rheinland-Pfälzische Technische Universität Kaiserslautern-Landau, Erwin-Schrödinger-Straße 46, 67663 Kaiserslautern, Germany.

[2]Experimentalphysik II, Institute for Physics, Augsburg University, D-86135 Augsburg, Germany.

[3]School of Science, Hunan University of Technology, Zhuzhou 412007, China.

[4]Physics Department, Faculty of Science, Al-Azhar University, Nasr City, E-11884 Cairo, Egypt.

[5]Physics Department, Faculty of Science, Galala University, New Galala City, Suez, 43511, Egypt

[6]Centro de Física de Materiales CSIC/UPV-EHU- Materials Physics Center, 20018, San Sebastian, Spain.

[7]IKERBASQUE, Basque Foundation for Science, Plaza Euskadi 5, 48009 Bilbao, Spain.

[8]Physics Department E20, TUM School of Natural Sciences, Technical University of Munich, James-Franck-Straße 1, D-85748, Garching, Germany.

[*]Email: lu.lyu@uni-a.de

[*]Email: Benjamin.stadtmueller@uni-a.de

**Abstract**

Quantum confinement of surface electrons in two-dimensional metal-organic porous networks offers a powerful route to engineer electronic states for emerging quantum and spintronic technologies at the molecular scale. To date, such confinement has been understood primarily in terms of lateral, two-dimensional surface potential landscapes. Here, we demonstrate the pivotal role of vertical variations in the surface potential on the quantum confinement of surface electrons. We investigate the confinement behavior of both Shockley surface state and image potential state electrons with their distinct vertical electron density distributions in an exemplary Cu-T4PT metal-organic porous network on Cu(111). Advanced spectroscopic techniques reveal a substantial band renormalization for the image state electrons, manifested as a significant increase in their effective mass, whereas the Shockley state electrons remain almost unchanged. Such striking divergence is attributed to the distinct three-dimensional potential landscape of the Cu-T4PT network with a strong repulsive potential at the vertical position of the molecular backbone and leaky channels underneath the Cu coordination spheres. Our findings demonstrate the essential role of vertical potential engineering in designing quantum confined states and thus advance the control of electronic properties and quantum phenomena on the nanoscale.

## Introduction

Quantum confinement of electrons and other (quasi-) particles within artificial nanostructures on surfaces provides an intriguing opportunity to explore the superior functionalities of quantum mechanics laws in condensed matter systems. Along these lines, the exploitation of phenomena such as quantum states coherence and coherent transport is of paramount significance for the realization of quantum technologies such as quantum computing on the solid-state platform. A key milestone in accessing the quantum world on surfaces was initially the atomic manipulation schemes involving scanning tunneling microscopy tips, which enabled the creation of tailor-made artificial nanostructures such as quantum corrals[1], one-dimensional gratings[2], and even two-dimensional lattices[3]. Pioneering studies thereby demonstrated quantized electronic states in functional quantum corrals[4] and striking quantum mirages[5]. However, atomic manipulation protocols reached their practical limitations when applied to larger, periodic nanoarchitectures composed of hundreds or even thousands of atoms or molecules. In contrast, surface self-assembly, for instance, of organic molecules and metal atoms adsorbed into two-dimensional metal-organic porous networks (2D-MOPNs) has emerged as an efficient alternative for constructing highly ordered and coupled quantum dot arrays[6]. The unique advantage of such 2D-MOPNs lies in their chemical tunability, which allows precise control over the size, shape, and periodicity of nanoporous architectures through molecular design and intercomponent interactions[7]. This tunable platform has enabled the observation of quantum confined states within network pores and the realization of engineered two-dimensional band structures arising from the controlled coupling between neighboring quantum dots on metal surfaces[8,9].

Despite considerable progress, fundamental challenges persist in fully harnessing quantum-confined states within functional nanostructures for applications in information and quantum technologies. A central simplification in understanding quantum confinement of surface electrons—one that we seek to revisit—is the widespread assumption of a purely 2D surface potential landscape[10], as schematically illustrated in Fig. 1a. In the conventional models, the potential confining the surface state $\psi_0$ (green) is treated as a uniform barrier $V_0$ at the molecular backbone. The coordinating metal centers contribute to weakly repulsive, or even attractive, local potential channels that introduce in-plane variations in electron confinement across the metal-organic network. These spatial variations in the plane of the metal-molecule network structure have been successfully modeled by semiempirical simulations using electron plane wave expansion (EPWE), which, in combination with experiments, have revealed the presence of so-called leaky channels at metal coordination sites[11]. However, such 2D models neglect vertical modulations of the potential due to the intrinsic structure of molecular orbitals and their hybridization with the underlying substrate[12]. This vertical interaction can significantly reshape the interfacial potential, underscoring the need for a more complete and realistic treatment of quantum confinement beyond idealized 2D approximations.

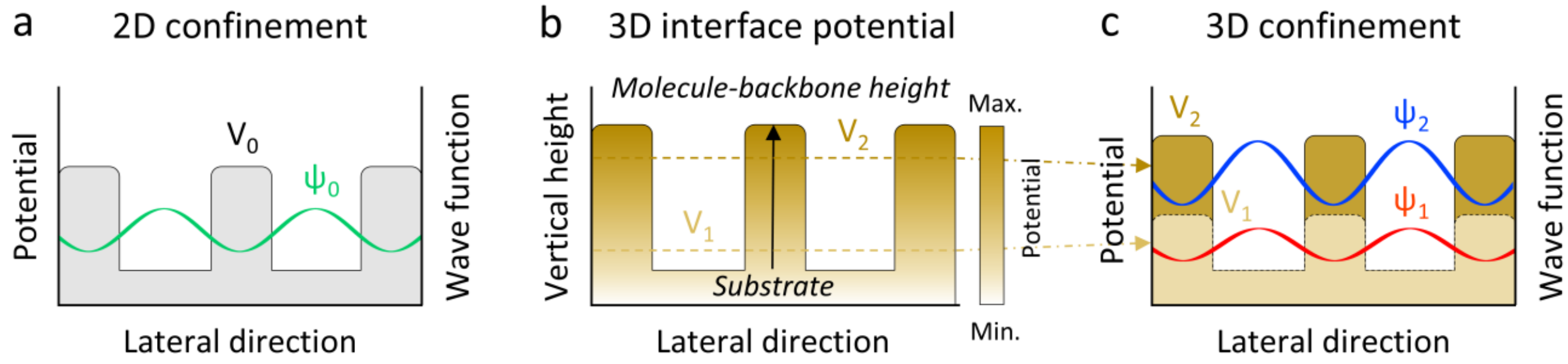

**Fig. 1 | Schematic illustration of different models for describing the quantum confinement of electrons in an organic porous network. a**, The currently established 2D confinement model considering a uniform lateral periodic potential $V_0$, neglecting the vertical potential variation. The confined electronic state $\psi_0$ (green) is located within the molecule-built potential wells. **b**, Model of a 3D interface potential showing an incremental vertical potential from the substrate to the molecule-backbone height (black arrow). Darker shading indicates higher potential. The light and dark brown dashed lines mark potential levels $V_1 < V_2$ at different heights. **c**, 3D confinement model incorporating both lateral and vertical potential variations. The vertical potential $V_1$ (light brown) and $V_2$ (dark brown), extracted from (**b**) (dashed arrows), produce a distinct vertical confinement for the electronic states $\psi_1$ (red) and $\psi_2$ (blue), respectively.

Interestingly, the crucial role of the third spatial dimension has not been uncovered so far as all existing studies have focused exclusively on the quantum confinement within the 2D free electron gas, associated with the so-called Shockley surface state (SS) electrons of noble metal surfaces. The vertical distribution of these surface electrons with respect to the surface is described by their wave function or more precisely, by their probability density given by the absolute square of the wave function. For typical noble metals such as Cu, Ag or Au, the maximum probability density of the SS is localized within the first atomic layer of the surface[13] that is substantially lower than the adsorption height of molecular adsorbates or adatoms. Lateral confinement of these surface electrons within a 2D-MOPN can be modelled quantitatively by adjusting the parameters of the simplified surface potential model. However, the confinement for SS is relatively weak, and the effective mass enhancement of the electrons is generally less than 60%[9,14]. As a consequence, the contributions of molecular backbones and coordinating metal atoms to confinement potential are significantly underestimated in the idealized 2D model.

Importantly, the Shockley surface state is not the only type of surface state that can be found on noble metal surfaces. Most famously, the surfaces can also host a Rydberg-like series of image potential states (IPSs). Like SS electrons, IPSs are free-electron-like states, but they are bound to the metal surface by the interaction of an excited electron just outside the metal surface and its induced image charge inside the substrate. This interaction leads to a sequence of unoccupied IPSs with a characteristic probability density maximum located approximately 3-5 Å above the surface[15], in striking contrast to the near-surface localization of the SS electrons. Recently, the confinement of IPSs has also been modeled through the utilization of 2D surface potentials[16].

To incorporate the effects of the modulation of the vertical potential, a three-dimensional (3D) interface potential model, as illustrated in Fig. 1b, more accurately describes the spatial confinement of surface electrons in the 2D-MOPN on a metal substrate. In this model, the barrier potential of the molecular backbone is partially screened by the electron density of the underlying metallic substrate, resulting in a weaker barrier potential closer to the substrate. Consequently, the two vertical potential $V_1$ and $V_2$, located at different heights (dashed lines in Fig. 1b), produce distinct confinement for the surface electronic states $\psi_1$ (red) and $\psi_2$ (blue), as illustrated in Fig. 1c. The simultaneous presence of SS and IPS, each exhibiting distinct vertical wave function distributions, thus provides a unique platform to explore the role of the third spatial dimension in surface states quantum confinement.

In this work, we quantify the quantum confinement of both the SS and the first IPS (n = 1) within a Cu-T4PT (2,4,6-tri(4-pyridyl)-1,3,5-triazine) porous network on the (111) oriented copper surface. Using near-threshold and two-photon photoemission (2PPE) experiments in combination with momentum microscopy[17,18], we uncover a substantially different band structure for the SS and the IPS, indicative of fundamentally distinct quantum confinement behaviors. These experimental results are corroborated by density functional theory (DFT) calculations, which expose a pronounced contrast in the potential landscape at the vertical positions of the probability density maxima of the two surface states. This clear divergence underscores the crucial role of the vertical variation of the scattering potential landscape in the engineering of quantum confinement at surfaces.

## Results

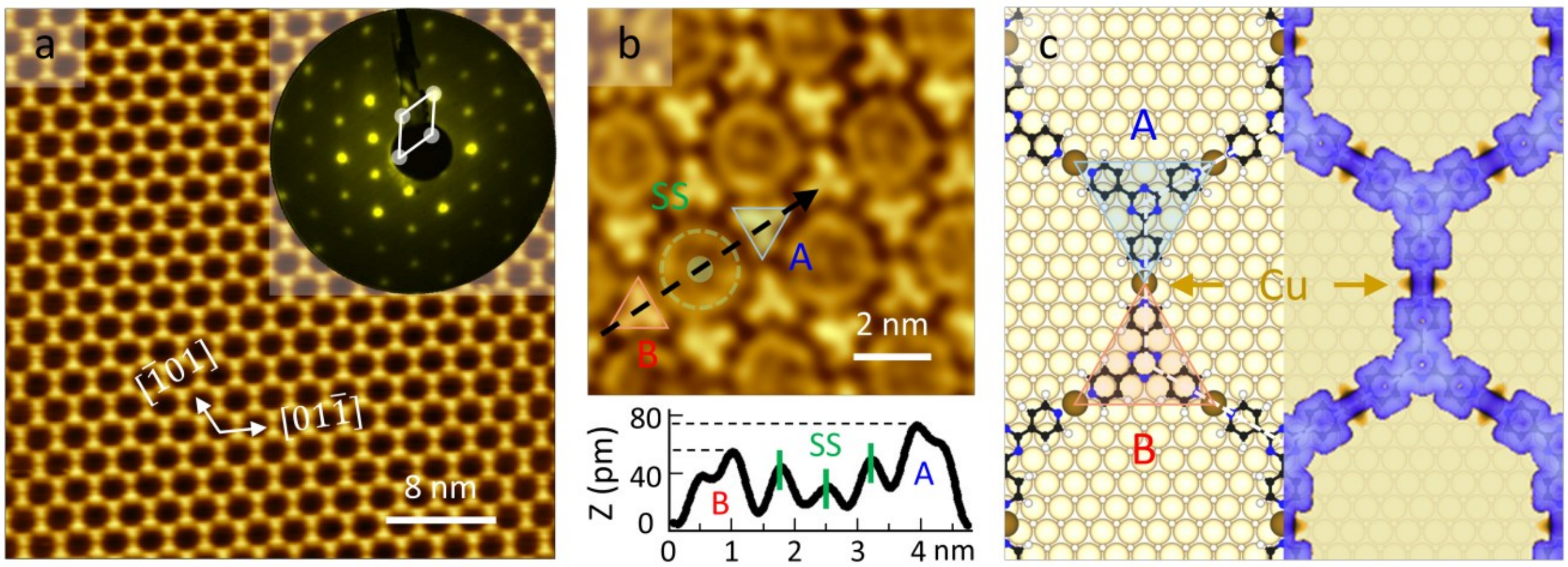

**Fig. 2 | Structure of the Cu-T4PT porous network on Cu(111). a**, Large-scale STM image ($V_{tip}$ = +1.00 V) shows a well-ordered porous superlattice aligned along the surface $[\bar{1}01]$ and $[01\bar{1}]$ directions. The inset displays the corresponding LEED pattern ($E_{kin}$ = 8 eV, inset), revealing a single-domain structure as indicated by the reciprocal-lattice spots (white). **b**, The molecular-resolution STM image (upper panel, $V_{tip}$ = +0.55 V, i.e. for occupied states) shows central protrusion and a surrounding ring-like feature of the spatial confinement of Shockley surface state (SS) electrons in the pores of the network structure. A line profile (lower panel) along the dashed arrow crosses the SS features (green dashed circle and center spot) and T4PT molecules labeled as A and B (light blue and red triangles), showing a distinct height difference between A and B. **c**, Simulation-optimized structural model (left side) of the Cu-T4PT network: molecules A and B coordinated with a Cu metal atom. The corresponding simulated STM image (right side) is calculated for a constant current STM experiment at 0.55 eV below the Fermi level, with an isosuface density of 1.5 e·Å$^{-3}$. All STM data were conducted at 106 K.

**The nanoporous architecture of the Cu-coordinated T4PT network.**

We begin by examining the structural and local electronic characteristics of the Cu-T4PT network. On the Cu(111) surface, the nitrogen ligands of T4PT molecules, bearing lone-pair $sp^2$ hybridized orbitals, spontaneously coordinate with native surface Cu atoms, leading to the self-assembly of a Cu-T4PT network[19]. The large-scale scanning tunneling microscopy (STM) and low-energy electron diffraction (LEED) measurements (Fig. 2a) highlight the formation of an almost perfect porous network with long-range order, exhibiting a $10 \times 10$ superlattice with a periodicity of $25.5 \pm 0.2$ Å along the $[\bar{1}01]$ and $[01\bar{1}]$ substrate directions. A molecular-resolution STM image (tip voltage bias $V_{tip} = +0.55$ eV, Fig. 2b) resolves the subtle structure of the pore. Each pore (~ 25 Å in diameter), surrounded by six T4PT molecules, displays both a central protrusion and a surrounding ring-like feature (green marks in the upper panel). These signatures at such a low bias suggest the presence of in-pore standing waves arising from confined Shockley surface electrons scattered by the potential landscape of the Cu-T4PT network. These confined states manifest as well-defined peaks in the corresponding line profile (lower panel). In addition, a distinct height difference in the local electronic density is observed between the two inequivalent T4PT molecules (labeled A and B), and are consistent with prior studies[19]. This interpretation is supported by density functional theory (DFT) calculations yielding the optimized Cu-T4PT network structure, shown in the left half of Fig. 2c. These calculations indicate that the network forms a mixed honeycomb-Kagome lattice according to the arrangements of the molecules and coordinating atoms. The neighboring molecules A and B, coordinated each with a Cu atom in a three-fold symmetry, adsorb at different sites of the substrate (details in Supplementary Fig. 1). Furthermore, the DFT-based STM topography simulation (Fig. 2c, right half) shows an almost undetectable feature for the coordinated Cu center, indicating a low local electron density that aligns well with the experimental image in Fig. 2b.

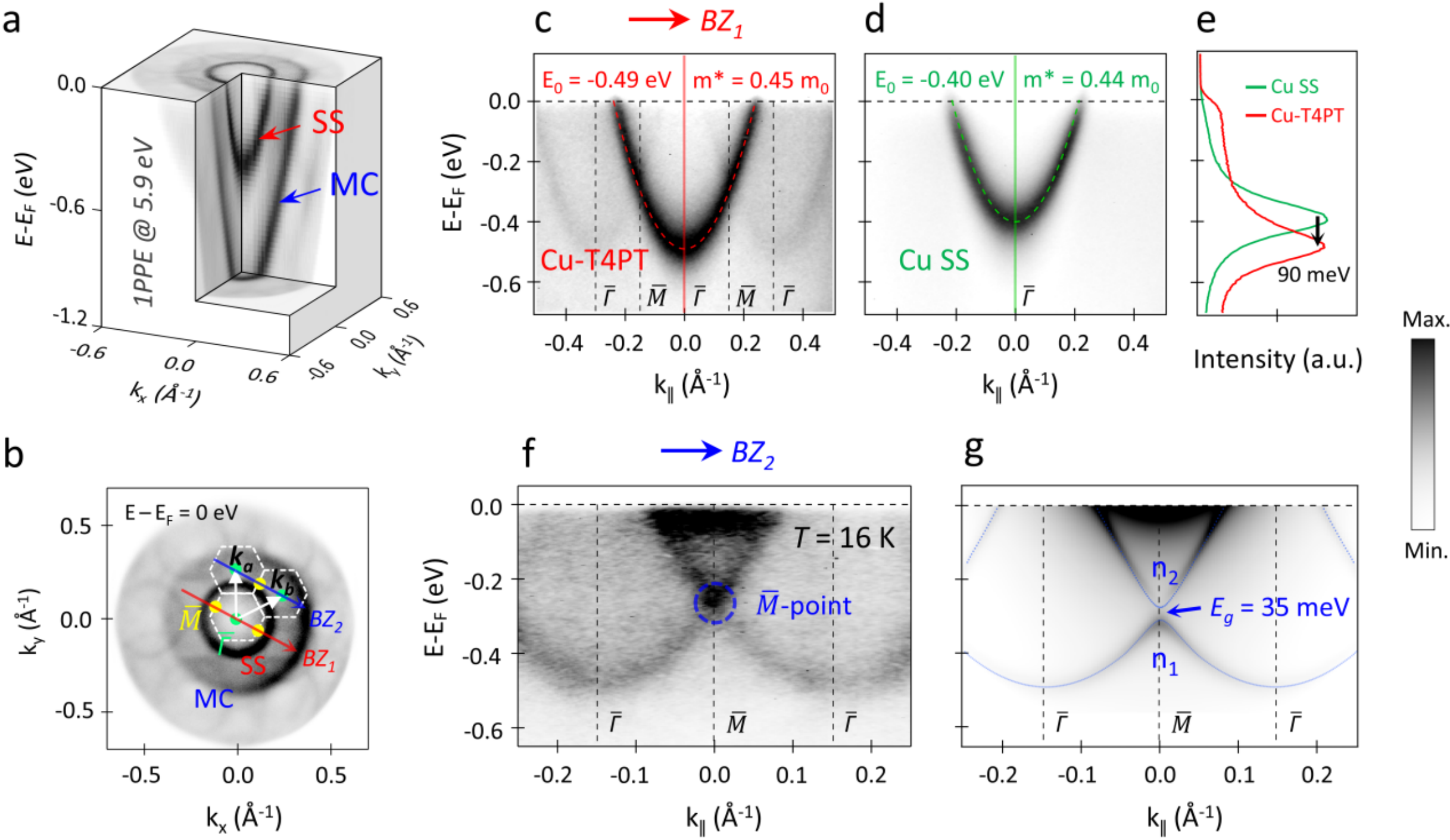

**Fig. 3 | Band dispersion of the Shockley surface state in the Cu-T4PT porous network. a**, Three-dimensional photoemission intensity in energy-momentum space of the Cu–T4PT monolayer network on Cu(111), acquired using the one-photon photoemission regime under grazing incidence illumination. The red and blue arrows indicate the Shockley surface state (SS) and the Mahan cone structure (MC), respectively. **b**, Constant energy map extracted from **a** at the Fermi level ($E - E_F = 0$ eV) revealing periodic replicas of the SS , with a characteristic reciprocal lattice vector of $|\boldsymbol{k_a}| = |\boldsymbol{k_b}| = 0.28 \pm 0.01$ Å$^{-1}$. The white dashed hexagons delineate the first and second Brillouin zones (BZs) of the Cu-T4PT network structure. High-symmetry points $\bar{\Gamma}$ and $\bar{M}$ are marked with yellow and light green dots, respectively. **c**-**d**, Angle-resolved photoemission spectroscopy (ARPES) data measured along the red arrow in **b** (BZ$_1$) for the Cu-T4PT network, and bare Cu(111) surface, respectively. The band parameters of $E_0$ (band bottom energy) and $m^*$ (electron effective mass) are extracted from fitting the surface state intensity with a parabolic dispersion (red and green dashed curves). **e**, Energy-distribution curves extracted at the $\bar{\Gamma}$-point (red and green lines) in (**c**) and (**d**), highlighting a 90 meV downward shift of $E_0$ upon the formation of the Cu-T4PT network. **f**, ARPES map recorded at $T$ = 16 K along the blue arrow in **b** (BZ$_2$) for the Cu-T4PT network, revealing a distinct contrast in intensity above and below the marked $\bar{M}$-point. **g**, Electron plane wave expansion (EPWE) simulations reproduce the observed SS bands in (**f**) and predict a bandgap of $E_g$ = 35 meV, with quantized states $n_1$ and $n_2$ above and below the gap-point. Data in (**a**, **b**) were recorded with $p$-polarized laser light and $h\nu$ = 5.9 eV. Data in (**c**-**f**) were acquired using a monochromatized VUV light source with $h\nu$ = 21.2 eV. (**a**-**e**) were measured at room temperature.

**Shockley surface state confinement in the Cu-T4PT porous network.**

The quantum confinement of both the SS and IPS in Cu-T4PT porous network, each exhibiting distinct vertical distributions, can be quantified by accessing their band dispersions using angle-resolved photoemission spectroscopy (ARPES) and momentum microscopy. In particular, the energy band shift, the electron effective mass, and the presence of superlattice gaps at the Brillouin zone boundaries serve as characteristic signatures of quantum confinement. These experimentally accessible quantities allow to gain insight into the confinement strength of surface states in a 2D-MOPN[8,14].

We first focus on the confinement of the occupied SS, which can be addressed in near-threshold one-photon photoemission (1PPE) with a photon energy of 5.9 eV. Fig. 3a presents the three-dimensional momentum microscopy datacube ($k_x$, $k_y$, E), in which two features with distinct parabolic dispersion are observed. The upper parabola, located near the Fermi energy ($E - E_F = 0$ eV), corresponds to the SS. The second feature at a larger binding energy arises from the so-called *sp*-*sp* Mahan cone (MC) transition, a well-known resonant bulk transition occurring only in the near-threshold photoemission regime for metals[20]. This feature has been reported on extensively[21-24] and does not play a further role for our discussion about surface electron confinement.

More importantly, the momentum-resolved photoemission experimental data reveals clear signatures of the SS confinement within the Cu-T4PT porous network, in full agreement with our STM results shown in Fig. 2b. The constant energy map (CEM, Fig. 3b) at the Fermi energy displays periodic replicas of the ring-like SS emission pattern, arranged with a reciprocal lattice periodicity of $|\boldsymbol{k_a}| = |\boldsymbol{k_b}| = 0.28 \pm 0.01$ Å$^{-1}$. This periodicity matches the size of the surface Brillouin zone (BZ) defined by the Cu-T4PT superlattice. In other words, the centers of these SS replicas are localized at the $\bar{\Gamma}$-points of the superlattice BZs (marked by white hexagons).

A more quantitative picture of the quantum confinement can be obtained by analyzing the band dispersion of the confined SS. For this, we conducted additional ARPES experiments with a higher resolution. Instead of a broadband laser source, we used a monochromatized vacuum ultraviolet VUV excitation source ($h\nu = 21.2$ eV), and the corresponding data is shown in Figs. 3c-f. It is well established that the SS electrons behave as quasi-free electrons parallel to the surface, giving rise to a parabolic dispersion[25] described by $E(k_{\parallel}) - E_F = E_0 + \frac{\hbar^2 k_{\parallel}^2}{2m^*}$, where $E_F$, $E_0$, $m^*$ and $k_{||}$ denote the Fermi energy, band minimum energy, effective mass, and momentum parallel wave-vector, respectively.

The ARPES intensity map with an energy-momentum relation in Fig. 3c displays the band structure of the confined SS for the Cu-T4PT network. The red arrow in Fig. 3b ($BZ_1$) marks the corresponding measurement location and direction in the momentum space. It reveals a series of band replicas along the $\overline{\Gamma M}$-direction across the adjacent first and second BZs. A parabolic fit to the central band yields an effective mass of $m^* = 0.45m_0$ ($m_0$ is the free electron mass), which is almost identical to the reference value of $0.44m_0$ for the clean Cu(111) surface (Fig. 3d). Interestingly, the energy distribution curves (EDCs, Fig. 3e) reveal a downward shift of approximately 90 meV for the band bottom $E_0$ for the Cu-T4PT network, in contrast to the typical upward shift associated with adsorption-induced Pauli repulsion (the push-back effect)[9,26]. Such an unusual energy shift likely originates from an attractive potential induced by the coordinated Cu atoms, which results in a hybridization with the confined SS[27].

The most crucial spectroscopic signature of quantum confinement-induced band renormalization is the opening of a weak superlattice band gap ($E_g$) at the BZ boundary, which can be quantitatively assessed from ARPES measurements. However, the negligible increase in $m^*$ suggests that $E_g$ is small, indicating a relatively weak confinement effect exerted by the Cu-T4PT network. Moreover, the very strong photoemission intensity of the SS in the first BZ (high photoemission cross section) obscures any visible gap signatures near the $\overline{M}$-points in Fig. 3c. To circumvent these limitations, we recorded an additional ARPES dataset at a low temperature ($T$ = 16 K) for the Cu-T4PT network, along a momentum cut through two adjacent second BZs, as indicated by the blue arrow in Fig. 3b ($BZ_2$). The ARPES map in Fig. 3f reveals the dispersion of two SS bands in neighboring BZs with almost identical spectral yield. Importantly, both bands display a characteristic intensity contrast at the crossing $\overline{M}$-point above and below the energy of $E - E_F = -0.30$ eV, providing the evidence of a small $E_g$ at the BZ boundary. This experimental observation is supported by EPWE simulations shown in Fig. 3g. The calculated SS dispersion using a 2D surface potential model of the Cu-T4PT reproduces the key features of the experimental data, corroborating a band gap of $E_g$ = 35 meV at the $\overline{M}$-point. Further EPWE simulations (Supplementary Fig. 2) indicate that the discrete SS eigenstates on either side of the gap exhibit distinct spatial density distributions within the nanopores, consistent with the standing-wave patterns observed in STM (Fig. 2b).

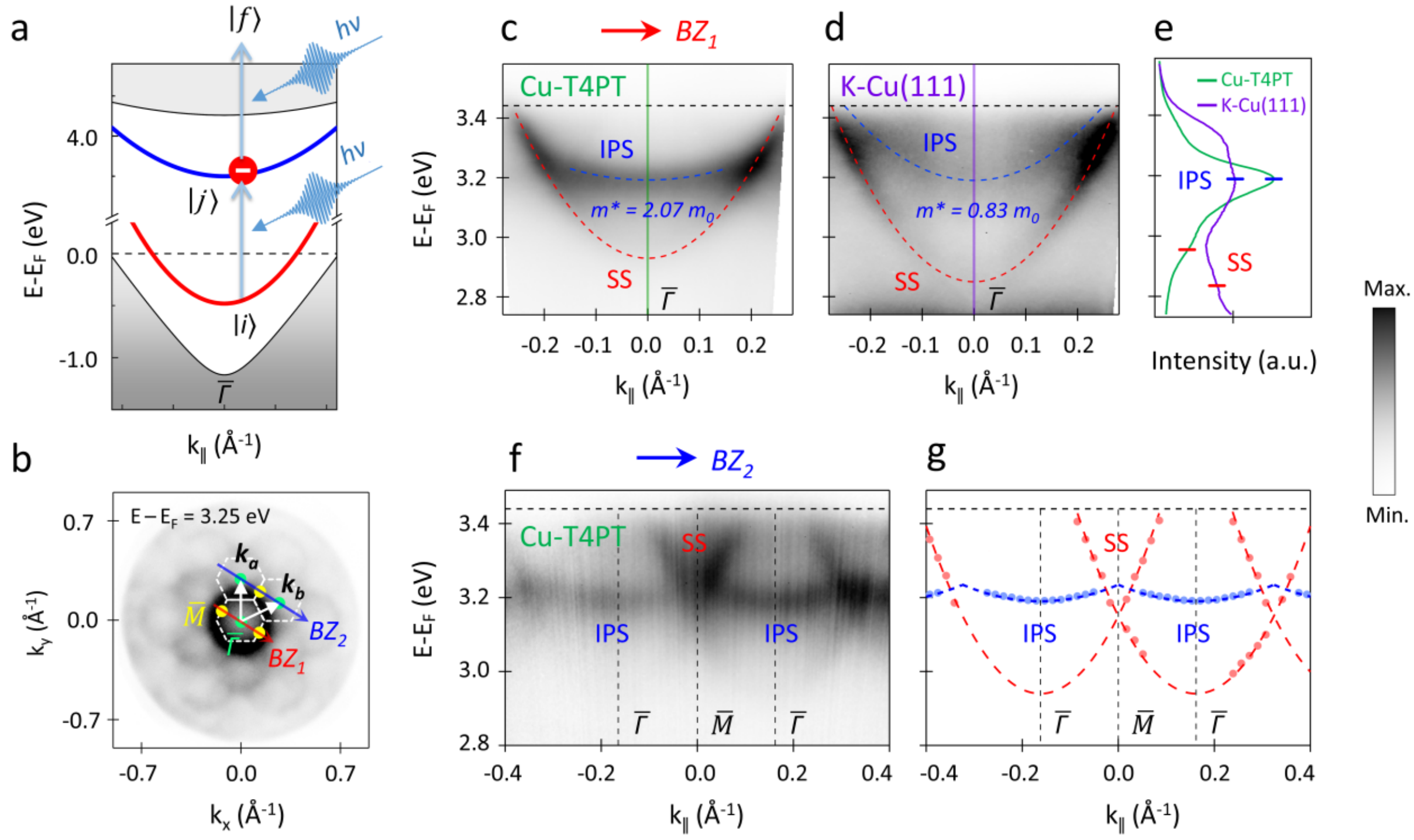

**Fig. 4 | Band structure of the first image-potential state (n = 1, IPS) in the Cu-T4PT porous network. a**, Schematic illustration of the one-color two-photon photoemission (2PPE) process used to probe unoccupied states above the Fermi level. One photon of the pulse excites electrons from an occupied initial state $|i\rangle$ into an unoccupied intermediate state $|j\rangle$, which is further promoted to the final state $|f\rangle$ above the vacuum level via a subsequent absorption of another photon of the same pulse. **b**, Constant energy map at an intermediate state energy of $E - E_F = 3.25$ eV, showing ring-like replicas of the unoccupied first IPS bands with a reciprocal lattice periodicity of $|\boldsymbol{k_a}| = |\boldsymbol{k_b}| = 0.28 \pm 0.02$ Å$^{-1}$. The white dashed hexagons mark the first and second Brillouin zones (BZs) of the Cu-T4PT lattice. High-symmetry points $\bar{\Gamma}$ and $\bar{M}$ are marked with yellow and light green dots, respectively. **c**-**d**, ARPES band dispersions recorded along the red arrow ($BZ_1$) for the Cu-T4PT network, and for the Cu(111) surface doped with a fraction of a monolayer (0.3ML) of potassium (K-Cu(111). The blue and red parabolas indicate the fit of the band dispersion for the IPS and the SS, respectively. **e**, Energy-distribution curves (EDCs) extracted at the $\bar{\Gamma}$-point (green and purple lines) from (**c**) and (**d**), highlighting an identical IPS energy (band bottom positions) of $E_0 - E_F = 3.18$ eV. **f**, ARPES intensity map measured along the blue arrow in (**b**) ($BZ_2$) for the Cu-T4PT network, showing two practically flat IPS bands from adjacent second BZs that intersect at the $\bar{M}$-point. **g**, Extracted band positions (solid dots) from (**f**), overlaid with parabolic fits for the IPS (blue) and SS (red) bands. All measurements at room temperature using a *p*-polarized laser with a photon energy of $h\nu = 3.44$ eV.

**First image potential state (n = 1, IPS) confinement in the Cu-T4PT porous network.**

We now turn to the quantum confinement of the IPS electrons in the Cu-T4PT network, which exhibit substantially different vertical probability density distributions compared to the SS electrons. Owing to their excited-state character, IPSs cannot be probed in the same linear photoemission experiment as the SS. Instead, we employ two-photon photoemission (2PPE) experiments, using the second harmonic of a titanium-sapphire (Ti:Sa) laser oscillator with a photon energy of 3.44 eV. In this approach, absorption of a photon from the laser pulse excites an electron from the occupied states into the intermediate IPS. A second photon from the same pulse subsequently photoionizes the electron into the vacuum as illustrated in Fig. 4a. By resolving the energy and momentum of the emitted photoelectrons, we obtain momentum space signatures of the initial state ($|i\rangle$), intermediate state ($|j\rangle$) and final state ($|f\rangle$) simultaneously, enabling a quantitative measurement of the band dispersion of the IPSs.

The corresponding 2PPE measurements are presented in Fig. 4b-g. The CEM in Fig. 4b, taken near the vacuum level ($E - E_F = 3.25$ eV), reveals again clear ring-like features indicative of periodic replicas of the IPS emission pattern in each surface BZ of the Cu-T4PT superlattice. This is very similar to the replicas of the SS observed in Fig. 3b, and thus points to a quantum confinement of the excited IPS electrons. A more quantitative view onto the IPS band dispersion can again be obtained with additional ARPES experiments in the 2PPE regime. Similar to SS electrons, IPS electrons exhibit a free-electron-like parabolic dispersion near the vacuum level, which follows a Rydberg-like series: $E - E_F = E_{vac} - \frac{0.85\ \text{eV}}{(n+a)^2} + \frac{\hbar^2 k_\parallel^2}{2m^*}$, where $E_{vac}$ is the vacuum level energy, $n$ is the image state index, and $a$ is the quantum defect parameter related to the bulk band structure[28]. The ARPES intensity map in Fig. 4c, recorded along the red arrow in Fig. 4b ($BZ_1$), clearly shows two parabolic bands with substantially different curvatures corresponding to the SS and the first IPS. The SS band matches the dispersion observed in the linear photoemission experiment (Fig. 3c). It is visible in our 2PPE data due to a direct two-photo-ionization process via virtual transition from the occupied surface state to the vacuum[29]. In contrast, the IPS band appears as very flat with its band bottom energy ($E_0$) located 3.18 eV relative to $E_F$, despite being a free electron-like surface state. A parabolic fit to the IPS band yields an effective mass of $m^*_{IPS_Cu-T4PT}$ = 2.07 $m_0$, which is approximately 2.6 times larger than the value for the pristine Cu(111) ($m^*_{IPS_Cu}$

= 0.81 $m_0$, see Supplementary Fig. 3). Moreover, the IPS band bottom $E_0$ shifts towards $E_F$ by 0.77 eV compared to the clean Cu(111) and is now located deeper in the L-band gap of the substrate.

This exceptionally large increase in the effective mass of the IPS band provides a direct signature of strong quantum confinement of the IPS electrons within the 3D interface potential landscape of the Cu-T4PT network and not caused by the adsorption-induced band shift of the IPS. To further support this conclusion, we monitored the effective mass of the IPS on clean Cu(111) at different band bottom positions through doping the surface with potassium (K) atoms. Here, gentle charge doping induces a continuous downward shift of the IPS energy due to the lowering of the surface work function. Upon deposition of 0.3 monolayers (ML) of K atoms, the IPS band on the K-doped Cu(111) surface in Fig. 4d shifts to the same energy position as the one observed for the Cu–T4PT network, as confirmed by the additional EDC curves (Fig. 4e). However, its effective mass is only $m^*_{IPS}$ = 0.83$m_0$ and thus identical to the one of the undoped surface. This observation unambiguously demonstrates a negligible change in the coupling of the IPS state with the Cu bulk bands upon lowering the IPS into the L-band gap. Further insight is provided by additional ARPES measurements on the Cu-T4PT network along the direction marked with a blue arrow ($BZ_2$) in Fig. 4b. The map, shown in Fig. 4f, reveals a remarkably flat first IPS band entirely residing within the Cu bulk band gap. The extracted band positions (method in Supplementary Fig. 4) plotted in Fig. 4g clearly demonstrate that the first IPS band in the Cu-T4PT network has a periodic and quantized dispersion structure.

In this regard, our experimental results not only show that optically excited IPS electron can also be confined in a 2D-MOPN, but importantly, the comparative analysis between SS and IPS confinement within the same material system highlights a striking dependence of quantum confinement strength on the vertical wavefunction distribution of the surface state. This underlines the crucial role of the vertical component of the 3D interface potential landscape in governing the quantum confinement of surface state electrons.

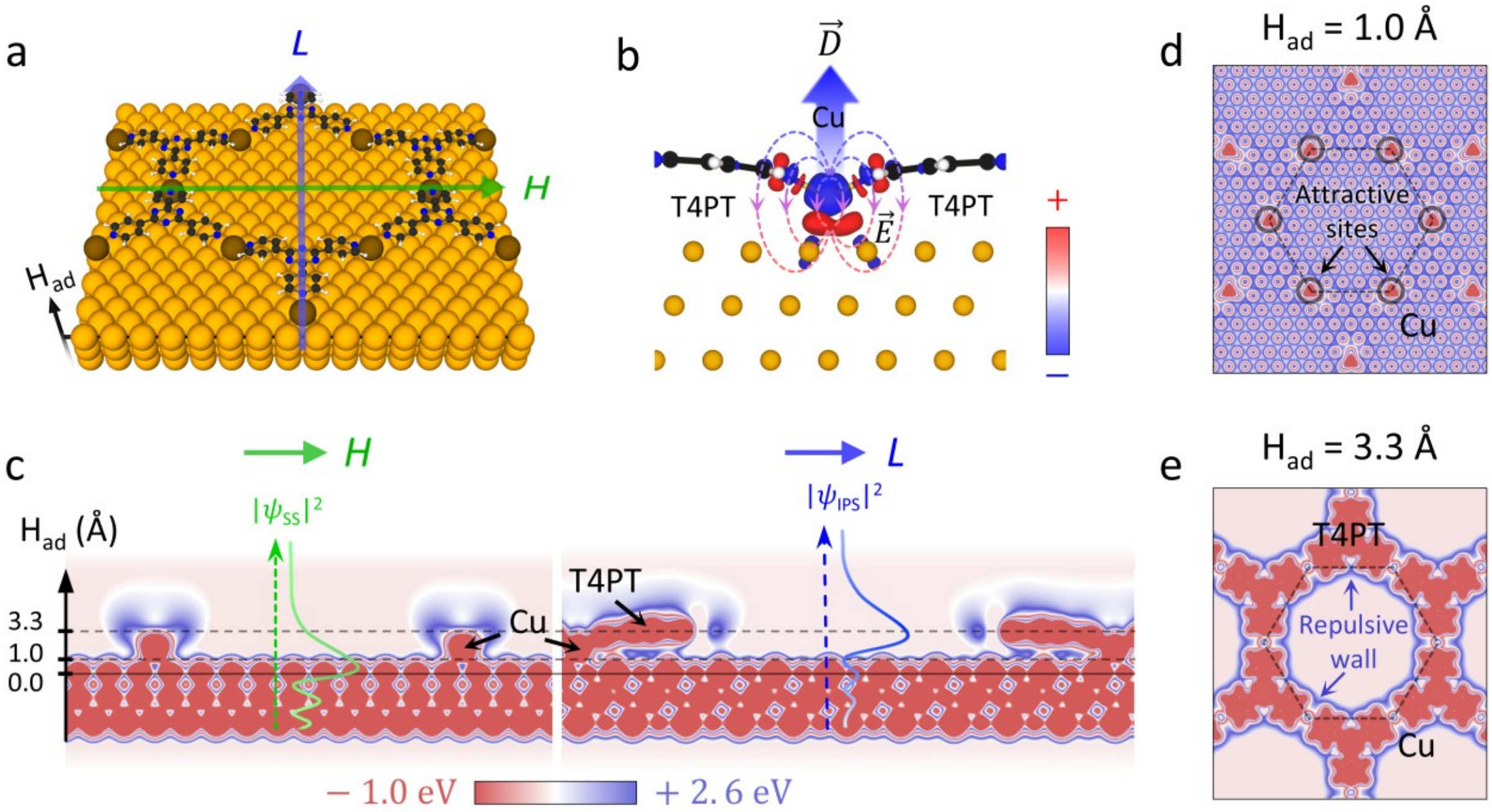

**Fig. 5 | Three-dimensional potential landscape in the Cu-T4PT porous network. a**, Top view of the Cu-T4PT structure on the Cu(111) surface. The horizontal (*H*, green) and longitudinal (*L*, blue) directions correspond to axes across the coordinated Cu atoms (brown spheres) and T4PT molecules, respectively. **b**, Side view of the charge density difference plot at one Cu-T4PT coordination site. Red (charge accumulation) and blue (charge depletion) regions indicate downward pointing electric fields ($\vec{E}$, dashed lines) and an upward pointing interfacial dipole ($\vec{D}$, blue arrow). **c**, Simulated electrostatic potential (ESP) maps along cross-sections of the *H* (left) and the *L* (right) directions, as marked in (**a**). The vertical adsorption heights ($H_{ad}$) of 1.0 Å and 3.3 Å, measured from the top Cu layer ($H_{ad}$ = 0.0 Å), correspond to the average height of the Cu surface and T4PT backbones, respectively. The blue and red areas represent repulsive and attractive potential, respectively. Model curves within the pore illustrate the vertical probability distribution of the Cu(111) Shockley state (SS, green) and the first image-potential state (IPS, blue). **d**, Planar ESP map at $H_{ad}$ = 1.0 Å highlights attractive potential sites at coordinated Cu atoms. **e**, Planar ESP map at $H_{ad}$ = 3.3 Å reveals a sealed repulsive potential wall surrounding the Cu–T4PT pores. Isosurface values in (**b**): +/–0.004 e·Å$^{-3}$.

## Discussion

The microscopic origin of the strikingly contrasting quantum confinement behaviors for the Shockley SS and the IPS can be elucidated through ab-initio DFT calculations of the Cu-T4PT network (see Methods section for details). A fully relaxed and optimized model of the porous network is shown in Fig. 5a, in which each Cu adatom is coordinated by two adjacent T4PT molecules.

The local interactions among the molecules, the Cu atom and the substrate induce a characteristic charge redistribution at the Cu coordination site, which is clearly illustrated in the charge density difference plot in Fig. 5b. Notably, a distinct charge depletion is found at the position of the Cu adatoms (blue region), accompanied by a charge accumulation barely underneath (red region), which coincides with the formation of the downward electric fields ($\vec{E}$) and the upward dipole ($\vec{D}$) at the Cu site. These local surface (or interfacial) dipoles can significantly reduce the work function[30] and lower the vacuum level by 0.85 eV (see Supplementary Fig. 5). Consequently, the energy position of the IPS shifts deeper in the L-band gap of the Cu substrate, consistent with the experimentally observed energy shifts (more parameters summarized in Supplementary Table. 1).

The local charge redistribution, including the charge transfer processes within the porous network structure, reflects the interfacial hybridization and thereby reshapes the 3D interface potential landscape responsible for the confinement of both SS and IPS electrons. To figure out the impact of the vertical modulation of this potential, we evaluated the vertical electrostatic potential (ESP) landscapes (Fig. 5c) perpendicular to the Cu(111) surface, extracted along two representative directions ("*H*" and "*L*" marked in Fig. 5a): through the Cu coordination sites (left) and the molecular backbones (right). Overall, both potential landscapes exhibit a predominantly repulsive electron potential (blue regions) in both the T4PT molecules and Cu atoms, which is consistent with the existence of quantum wells for surface electrons in the Cu-T4PT network pores. However, on closer inspection, we find characteristics of strong repulsive potential residing at the vicinity of the T4PT molecule for the average adsorption height of its backbone, while the Cu sites only show a repulsive potential at large surface distances ($H_{ad}$ approx. 3 Å) and an attractive potential very close to the surface ($H_{ad}$ approx. 1 Å). These distinct vertical potential distributions are consistent with the schematic 3D interface potential model (Fig. 1b) and account for the strikingly different confinement strengths of the SS and the IPS electrons. We illustrate this by including the vertical probability density of the Cu(111) Shockley SS ($|\psi_{SS}|^2$) and first IPS ($|\psi_{IPS}|^2$) in Fig. 5c. The majority of the SS density is located very close to the metal surface (peak approx. 1 Å), where it overlaps with the attractive region near the Cu sites. This behavior is further illustrated by the 2D ESP map in Fig. 5d (extracted at a vertical height $H_{ad}$ = 1 Å). The attractive sites of the Cu atoms act as leaky channels for the SS electrons, weakening the lateral confinement and explaining the limited band renormalization observed in ARPES.

In contrast, the IPS electrons are predominantly located at a vertical position of approx. 3.3 Å, coinciding with the adsorption height of the T4PT molecules. At this height, the peripheries of the T4PT backbones and the coordinating Cu atoms form a sealed and almost ring-like repulsive potential wall around each pore as shown in the 2D ESP map (Fig. 5e). Consequently, the combination of a strongly repulsive potential at the adsorption height of the network components with the absence of any leaky channels at the same vertical position is responsible for the stronger confinement of the IPS electrons within the Cu-T4PT porous network, and thus for the formation of an almost flat IPS band by the quantum confinement.

In conclusion, our work demonstrates the pivotal role of the three-dimensional interface potential landscape for the quantum confinement of different types of surface electrons in porous nanostructures. By systematically examining both Shockley surface states (SS) and image potential states (IPS)—each with distinct vertical electron density distributions—we uncover striking disparities in band dispersion and effective mass renormalization within a Cu-coordinated T4PT porous network. These differences arise from vertical potential variations: leaky channels at Cu sites close to the surface result in weak confinement of the SS, whereas a strong, sealed repulsive potential at the adsorption height of the molecular backbones leads to a giant confinement of the IPS electrons.

Our findings establish a significantly expanded framework for engineering quantum confinement by harnessing the full three-dimensional complexity of surface potential landscapes. This advancement represents a critical leap forward, enabling the simultaneous and selective confinement of multiple surface states within individual nanopores. By achieving this level of control, we open the door to realizing coherent quantum interference and superposition of surface states in artificial nanostructures, key prerequisites for the development of robust quantum computing architectures and next-generation nanoscale quantum technologies.

## Methods

**Sample preparation.** All experiments were carried out in two commercial ultra-high-vacuum (UHV) systems (base pressure ~$10^{-10}$ mbar): a SPECS ARPES setup and an in-situ multi-chamber Omicron VT-STM, the latter integrated with a FOCUS PEEM chamber. The Cu(111) substrate (MaTecK) was cleaned through multiple cycles of Ar-ion sputtering (1.0–1.8 kV) and annealing (610–650 ℃). T4PT molecules (TCI, purity > 97%) were thermally evaporated at 450 K with a deposition rate of 0.33 ML·min$^{-1}$ (1 ML = full monolayer Cu-T4PT network on Cu(111)). After deposition, the T4PT film was annealed at 410 K for 30min to form a defect-free Cu-coordinated T4PT network. Potassium atoms (SAES Getters) were deposited at a rate of 0.07 ML·min$^{-1}$ on clean Cu(111). During all depositions, substrates were held at room temperature.

**Scanning tunneling microscopy (STM).** STM measurements were performed at 106 K (liquid nitrogen cooling) in constant current mode, using a tunneling current ($I_t$) of 70–90 pA. The bias voltage ($V_{tip}$) is applied to the STM tip, and the positive and negative values correspond to tunneling into the samples' occupied and unoccupied states, respectively. All STM images were processed using Gwyddion software[31].

**Momentum microscopy (MM).** Momentum-resolved photoemission measurements were performed using a photoemission electron microscope (FOCUS PEEM) combined with a time-of-fight delay-line detector (DLD). Three-dimensional datasets ($k_x$, $k_y$, $E_{kin}$) were acquired with energy and momentum resolutions better than $\Delta E < 100$ meV and $\Delta k < 2 \times 10^{-2}$ Å, respectively. The excitation sources included the third (THG, $h\nu$ = 4.6 eV) and fourth (FHG, $h\nu$ = 5.9 eV) harmonics of a commercial pulsed titanium-sapphire (Ti:Sa) laser (Tsunami, Spectra-Physics, central wavelength of 840 nm), with nonlinear conversion achieved using a THG Tripler-Box and a BBO crystal, respectively. For the two-photon photoemission (2PPE) experiments, the second harmonic (SHG, $h\nu$ = 3.44 eV) from a separate femtosecond Ti:Sa laser (Mai Tai, Spectra-Physics, central wavelength of 720 nm) was employed. The samples were illuminated with linearly *p*-polarized light under grazing incidence (65°) relative to the surface normal direction.

**Angle-resolved photoemission spectroscopy (ARPES).** ARPES spectra were acquired using a hemispherical energy analyzer (PHOIBOS 150, SPECS) equipped with a CCD detector. The energy and momentum resolutions were better than 5 meV and $1 \times 10^{-2}$ Å$^{-1}$, respectively. One-photon photoemission (1PPE) measurements were performed using a monochromatic He I$\alpha$ discharge lamp (21.2 eV, Scienta VUV 5K). For the two-photon photoemission (2PPE) experiments, we used the same laser source (Mai Tai, Spectra-Physics) and SHG at $h\nu$ = 3.44 eV, as mentioned above. The incidence angle of the *p*-polarized laser beam is 45° relative to the normal direction of the surface. A sample bias voltage of −4 V was applied to the measurements, and the distortion of low-kinetic photoelectrons was corrected[32] considering the sample-detector potential differences. The cooling measurement at 16 K was conducted by liquid helium.

**Electron Plane Wave Expansion (EPWE) simulations.** EPWE was performed using the code developed by Javier Garcia de Abajo, widely applied in the study of 2DEG scattering in periodic nanostructures with arbitrarily shaped barriers[6,33]. Within this code, muffin-tin potentials are assigned to the regions of molecule ($V_1$) and substrate ($V_0$). The wave function is expressed as a linear combination of plane waves, and the central form of the Schrödinger equation is solved to obtain the energy eigenvalues and wave functions across all momentum values. A 10 × 10 reciprocal lattice corresponding to 100 plane waves was used to ensure satisfactory convergence in the Fourier expansion of the potential. EPWE was used to calculate the band structures, density of states, and photoemission maps for surface states confinement within the Cu-T4PT porous network.

**Ab initio calculations.** Density functional theory (DFT) calculations were carried out with the VASP code[34]. The projected augmented wave (PAW) method[35] was employed with a plane-wave cut-off of 400 eV. The electron-exchange correlation was treated using the generalized gradient approximation (GGA) with the Predew-Burke-Ernzerhof (PBE) function[36]. Additionally, the nonlocal exchange-correlation function of optB86b+vdW[37] was included for van der Waals interactions. A supercell slab of the Cu-T4PT system contains three Cu(111) layers, two T4PT molecules, and three coordinated Cu atoms, with a vacuum gap of 15 Å in the z-direction. All atomic positions were relaxed with a force tolerance below 0.02 eV/Å. Only the Γ-point was sampled in the geometry optimization. STM simulations were based on the Tersoff-Hamann approximation[38], and the charge density contour was generated as constant-current images in P4VASP[39]. For the optimized Cu-T4PT structure, the spatial electrostatic potential (ESP) map was calculated using the VASP code, considering only the Hartree potential contributions. All structural visualizations were created using VESTA[40].

## Acknowledgements

This work was supported by the Deutsche Forschungsgemeinschaft (DFG, German Research Foundation), TRR 173-268565370 Spin + X: spin in its collective environment (Project B05 and B14). Scientific Research Fund of Hunan Provincial Education Department of China (Grant No. 21B0548, Grant No. 19B159). The National Natural Science Foundation of China (Grant No. 11704114).

## Author contributions

L.L, B.S. and M.A. conceived the experiments. L.L. performed the STM and LEED experiments, and L.L. analyzed the data. L.L., T.E., W.Y. carried out the ARPES and MM measurements and analyzed the results. J.X. performed the DFT calculations, and L.L. and J.X. discussed and interpreted the data. Z.M.A.E.-F., M.A., I.P.-Z. and J.V.B. conducted and analyzed the EPWE simulations. L.L. and B.S. wrote the manuscript with input from all authors.

## Supplementary information for:

# Exploring the third dimension in quantum confinement of surface electrons

### Supplementary Figures:

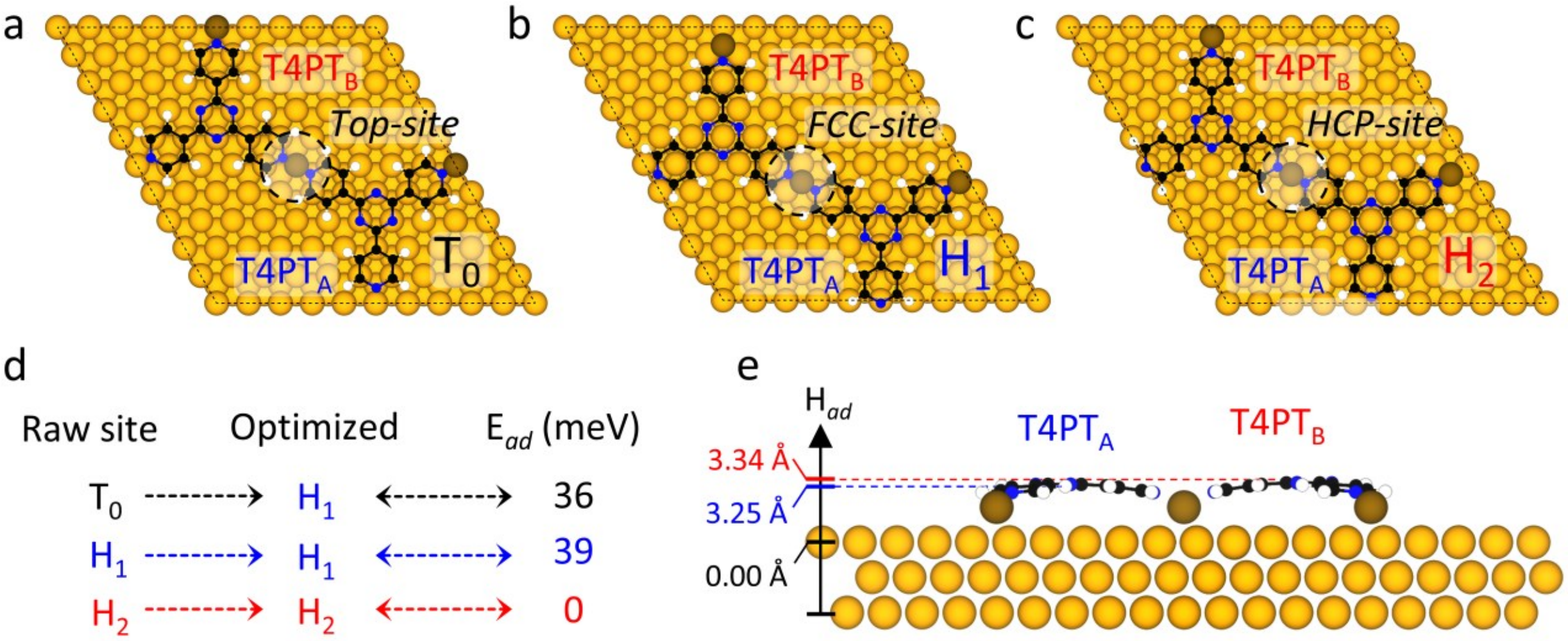

**Figure S1 | Structural optimization of Cu-T4PT coordination on Cu(111). a**-**c**, Three structural models of the Cu-T4PT network, defined by the adsorption site of the coordinated Cu atom (Coord-Cu): top site ($T_0$), FCC hollow site ($H_1$) and HCP hollow site ($H_2$). In each unit cell, two T4PT molecules ($T4PT_A$ and $T4PT_B$) are symmetrically placed but located at different adsorption sites. **d**, The adsorption energies ($E_{ad}$) of the three configurations, calculated as $E_{ad} = E_{Coord\text{-}Cu} + E_{T4PT_A} + E_{T4PT_B} + E_{Cu(111)} - E_{Cu\text{-}T4PT/Cu(111)}$, and $E_{ad}$ are relative to the $H_2$ structure. **e**, Side view of the optimized $H_1$ structure, showing the adsorption heights ($H_{ad}$) of the central backbones in $T4PT_a$ and $T4PT_b$, which are 3.25 Å and 3.34 Å, respectively.

The structural models of Cu-coordinated T4PT on Cu(111) are based on the two types of T4PT assemblies observed in STM experiments. In the three proposed structural models (Fig. S1a-c), the two T4PT molecules ($T4PT_A$ and $T4PT_B$) have distinct adsorption sites due to the Cu coordination (marked in the models). As shown in Fig. S1d, among the three configurations, the $H_1$ structure exhibits the largest adsorption energy ($E_{ad}$), indicating it is the most stable and preferred structure. A side view of the optimized $H_1$ geometry (Fig. S1e) reveals a small difference in adsorption height of $T4PT_A$ (3.25 Å) and $T4PT_B$ (3.34 Å). The adsorption heights are calculated from the average vertical distance between the central backbones (triazine unit) and the Cu(111) surface.

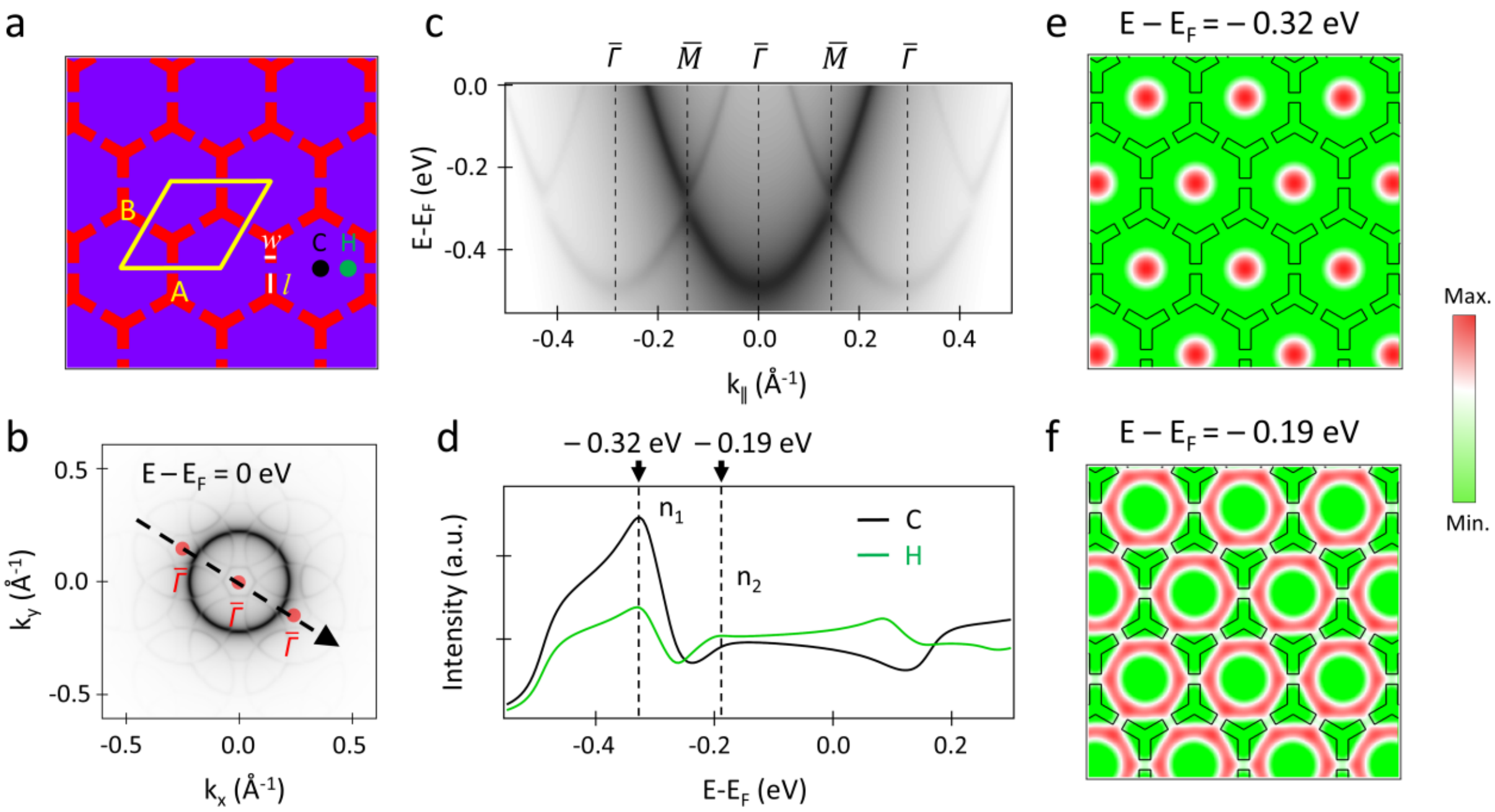

**Figure S2 | EPWE simulations of Shockley surface states (SS) in a Cu-T4PT model. a**, The 2D structural model of the Cu-T4PT network used for EPWE calculations, with lattice constant of |**A**| = |**B**| = 25.5 Å, molecular width ($w$) and length ($l$) of $w = l/\sqrt{3}$ = 3.0 Å, and molecular barrier potential of $V_1$ = 454 meV (red region). The coordinated Cu atoms and the substrate have the same potential $V_0$ = 0 meV (purple region). **b**, EPWE-simulated constant energy map (CEM) at $E - E_F$ = 0 eV, showing replicated patterns of the confined SS by the porous Cu-T4PT network. **c**, The simulated photoemission energy-momentum dispersion along the $k_{\parallel}$ direction (black dashed line in (**b**)). **d**, Simulated local density of states (LDOS) spectra at the center of a pore (C, black spot in (**a**)) and halfway (H, green spot in (**a**)), which indicate two resonance peaks of $n_1$ at –0.32 eV and $n_2$ at –0.19 eV, corresponding to different confined states. **e**, **f**, LDOS maps at the resonance energies, revealing $n_1$ as a dome-like state inside the pore and $n_2$ as a ring-like state surrounding it.

For the electron plane wave expansion (EPWE) simulations, we constructed the Cu–T4PT model (Fig. S2a) based on the experimental superlattice and molecular geometry, treating the coordinated Cu atom as a leaky channel with the same potential as the substrate ($V_0$ = 0 meV). The molecular region is defined by a barrier potential $V_1$ = 454 meV. The simulated constant energy map (CEM) at $E - E_F = 0$ eV (Fig. S2b) reproduces the replicated SS observed in experiments. The corresponding simulated energy-momentum dispersion (Fig. S2c) shows a small band gap at the $\bar{M}$ point. Due to lateral confinement, the LDOS spectra (Fig. S2d) display two distinct resonance peaks: $n_1$ at –0.32 eV and $n_2$ at –0.19 eV, corresponding to discrete SS eigenstates within the pore. The spatial LDOS maps (Fig. S2e,f) reveal that $n_1$ forms a dome-like distribution centered inside the pore, while $n_2$ forms a ring-like distribution encircling the pore. Their convolution is consistent with the standing-wave patterns observed in the STM measurements taken at the tip bias near-Fermi for the occupied region.

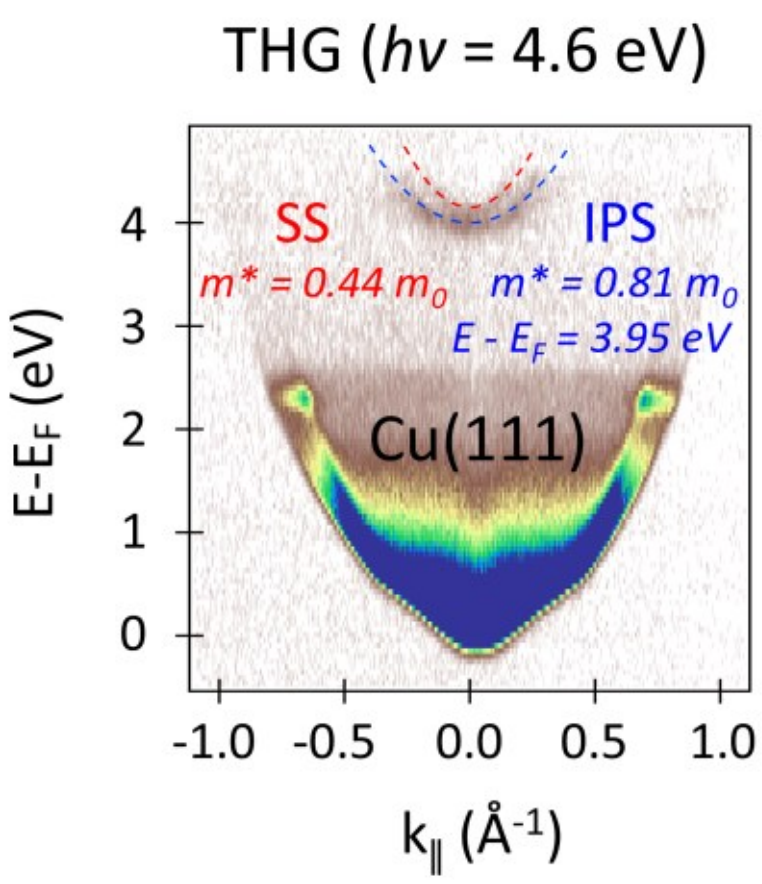

**Figure S3 | Two-photon photoemission (2PPE) experiments on Cu(111).** Energy distribution map of the bare Cu(111) measured with the third-harmonic (THG) of the laser at grazing incidence ($h\nu$ = 4.6 eV). The two fitted parabolas correspond to the first image potential state (IPS, blue) and the Shockley state (SS, red). The extracted effective mass of the IPS is $m^*$ = 0.81$m_0$, and the band bottom lies at $E_0 - E_F$ = 3.95 eV.

In Fig. S3, the 2PPE measurement with third harmonic of our laser shows the SS and first image potential state (n = 1, IPS) of the pristine Cu(111) surface. The SS signal originates from the occupied initial state, and the IPS signal is from the unoccupied intermediated state. From the fitted dispersion of the IPS, an effective mass of $m^*_{IPS}$ = 0.81$m_0$ is extracted, with a band minimum located at $E_0 - E_F$ = 3.95 eV.

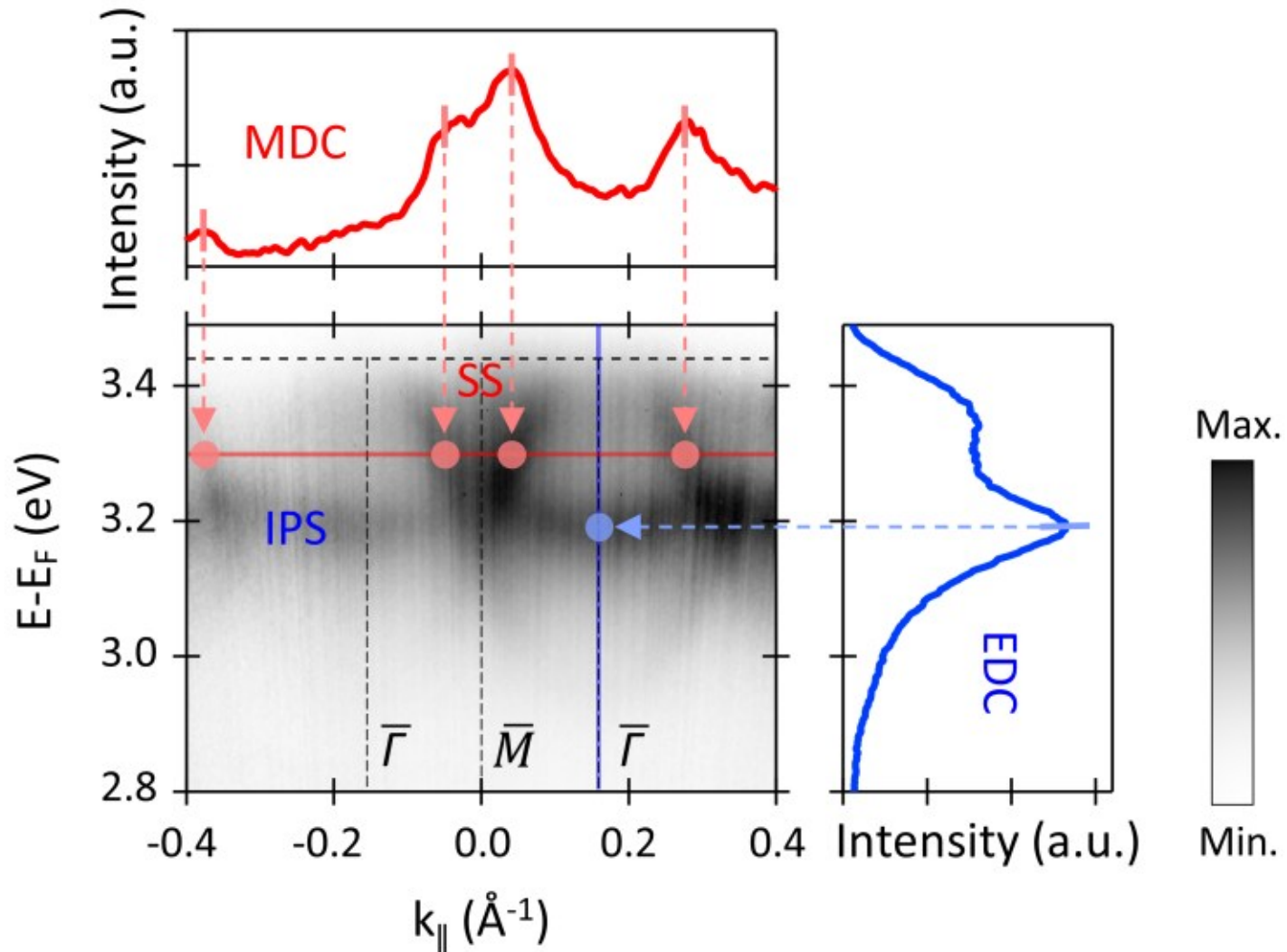

**Figure S4 | Extraction of band positions from the ARPES intensity map on the Cu-T4PT porous network.** ARPES intensity map measured along the $\bar{\Gamma}$-$\bar{M}$-$\bar{\Gamma}$ direction in the 2PPE measurement ($h\nu$ = 3.44 eV), showing the energy-momentum dispersions of the SS and IPS. Two representative line-profiles are shown: the blue one parallel to the energy axis ($E-E_F$) representing an energy distribution curve (EDC, right panel) at the $\bar{\Gamma}$-point ($k_\parallel$ = 0.14 Å$^{-1}$, Δk = 0.01 Å$^{-1}$), and red one parallel to the momentum axis ($k_\parallel$) representing an momentum distribution curve (MDC, top panel) at the $E - E_F$ = 3.30 eV (ΔE = 10 meV). Peak positions (light blue and red bars) extracted from the EDC and MDC indicate the band locations of the IPS (light blue spot) and the SS (light red spots), respectively.

Fig. S4 shows exemplary data to showcase the extraction of the band positions from the ARPES intensity map in Fig. 4f of the main text. The map captures the band dispersions of both the IPS and SS. The IPS shows a remarkably flat and periodic band along this direction. The vertical line-profile (blue), i.e., the energy distribution curve (EDC, right panel), taken parallel to the energy axis at the $\bar{\Gamma}$-point ($k_\parallel$ = 0.14 Å$^{-1}$), yields the IPS band positions (light blue spot) by identifying the energy value ($E - E_F$ = 3.19 eV) corresponding to the EDC peak. In contrast, the SS bands display strongly dispersive parabolic features. The horizontal line-profile (red), i.e., the momentum distribution curve (MDC, top panel), taken parallel to the momentum axis at $E - E_F$ = 3.30 eV, is used to locate the intersecting SS band positions (light red spots) by extracting the momentum values ($k_\parallel$ at –0.38, –0.05, 0.04 and 0.28 Å$^{-1}$) from the MDC peaks. By employing the EDC and MDC line-profiles, the positions of the remaining IPS and SS bands in the ARPES intensity map can be accurately identified, respectively.

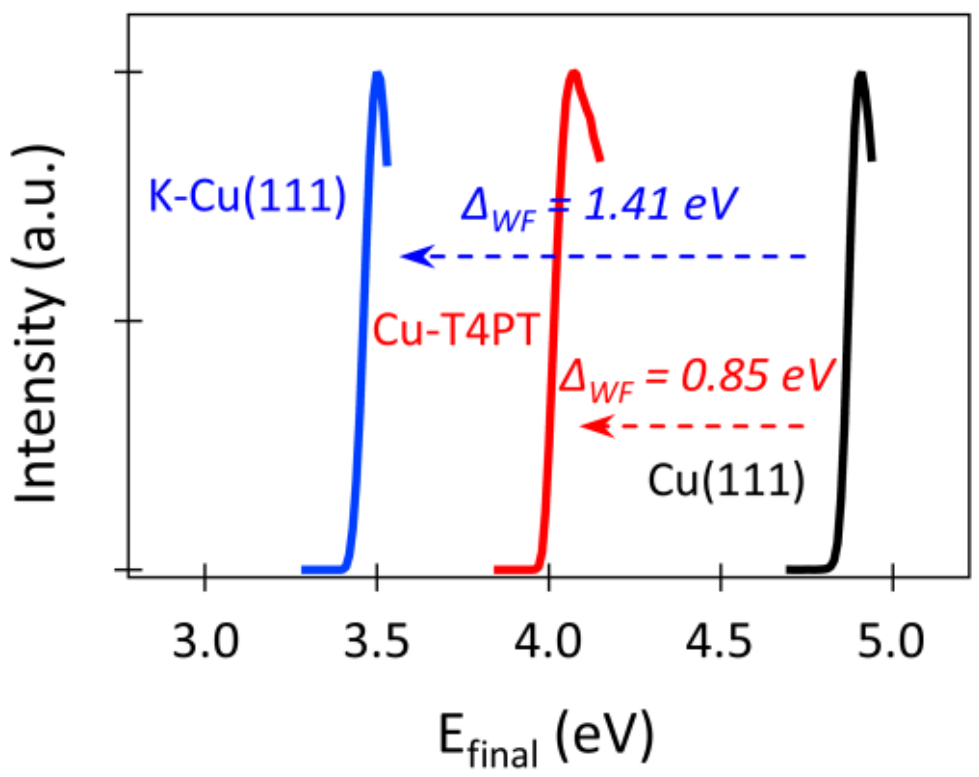

**Figure S5 | Evolution of the work function for Cu(111), Cu-T4PT network and potassium-doped Cu(111).** The secondary electron cutoff edges acquired along the $\overline{\Gamma}$ in the ARPES intensity map of the 2PPE measurements ($h\nu$ = 3.44 eV) reveal the work functions (WF) shifts. The WFs of Cu(111) (black), Cu-T4PT network (red) and potassium-doped Cu(111) (K-Cu(111), blue) are obtained by linear extrapolation of the lowest final state energies from the cutoff spectra. The WF decreases by 0.85 eV and 1.41 eV for the Cu-T4PT network and K-Cu(111), respectively.

Fig. S5 shows the secondary electron cutoff spectra used to determine the work functions (WFs) of pristine Cu(111), the Cu-T4PT network, and potassium-doped Cu(111) (K-Cu(111)). Upon the formation of the Cu–T4PT coordination network, an upward dipole moment emerges at the Cu-coordinated sites, reducing the surface potential and lowering the WF by 0.85 eV relative to bare Cu(111). For the K-Cu(111) system, the doped K atoms induce a strong charge transfer to the Cu surface, resulting in a more pronounced WF reduction of 1.41 eV.

**Table S1. Energy positions and effective masses of Shockley state and first image potential state in Cu(111), K-Cu(111) and Cu-T4PT network.** The table summarizes the band bottom energies and effective masses of the surface state (SS) and image potential state (IPS), as well as the work function (WF) for each system. $E_V - E_F$ is the energy differences between the vacuum level and the Fermi level, i.e., WF. ($E_0 - E_F$) and ($E_0 - E_v$) denote the SS and IPS bands bottom energies referring to the Fermi and vacuum level, respectively. $m^*/ m_0$ is the relative mass to a free electron. The last row reports the relative deviation of the effective mass from the clean Cu(111) reference, defined as $|(m^* - m^*_{Cu})/m^*_{Cu}|$, where $m^*_{Cu}$ is taken from the SS and IPS of the pristine Cu(111). The IPS effective mass in the Cu-T4PT system shows a pronounced increase, deviating by over 150% from the clean Cu(111) value, indicating a huge confinement effect by the porous Cu-T4PT network.

| | **Cu (111)** | | | **K-Cu(111)** | | | **Cu-T4PT** | | |
|---|---|---|---|---|---|---|---|---|---|
| | SS | IPS | WF | SS | IPS | WF | SS | IPS | WF |
| $E_V - E_F$ (eV) | \ | \ | 4.84 | \ | \ | 3.43 | \ | \ | 3.99 |
| $E_0 - E_F$ (eV) | −0.40 | 3.95 | \ | −0.59 | 3.18 | \ | −0.49 | 3.18 | \ |
| $E_0 - E_V$ (eV) | −5.24 | -0.89 | \ | −4.02 | −0.25 | \ | −4.48 | −0.81 | \ |
| $m^*/m_0$ | 0.44 | 0.81 | \ | 0.43 | 0.83 | \ | 0.45 | 2.07 | \ |
| $\|(m^* - m^*_{Cu})/m^*_{Cu}\|$ **(%)** | 0 | 0 | \ | 2.27 | 2.47 | \ | 2.27 | **156** | \ |